# Analysis and Forecasting of Fire incidence in Davao City

Merlito M. Villa[1] and Roel F. Ceballos[2]
[1] University of Southeastern Philippines, https://orcid.org/0000-0002-5830-5742
[2] University of Southeastern Philippines, https://orcid.org/0000-0001-8267-6482
[2] Email Correspondence: roel.ceballos@usep.edu.ph

## Abstract

Fire incidence is a big problem for every local government unit in the Philippines. The two most detrimental effects of fire incidence are economic loss and loss of life. To mitigate these losses, proper planning and implementation of control measures must be done. An essential aspect of planning and control measures is prediction of possible fire incidences. This study is conducted to analyze the historical data to create a forecasting model for the fire incidence in Davao City. Results of the analyses show that fire incidence has no trend or seasonality, and occurrences of fire are neither consistently increasing nor decreasing over time. Furthermore, the absence of seasonality in the data indicate that surge of fire incidence may occur at any time of the year. Therefore, fire prevention activities should be done all year round and not just during fire prevention month.

## 1.0 Introduction

Fire incidence can result in severe injuries, damage to personal property, and even death. It is a severe threat to life and property. Causes of fire incidence range from faulty wirings, discarded cigarettes on flammable materials, substandard products, and defective smoke detectors.

According to the World Health Organization (WHO, 2018) report, there are an estimated 180,000 global burn deaths every year, which usually occur in low-and middle-income countries. International Fire Statistics shows that 86% of all fire fatalities start in residential buildings, and 66% are caused by inappropriate human behavior (Burroughs, 2016). In the Philippines, 96,447 fire incidences have been reported nationwide from 2011 to 2017, causing 1,924 deaths, 5,750 injuries, and ₱31.06 billion in property loss (Commission on Audit, 2018). In Davao City, the Bureau of Fire Protection has reported a 35 percent increase in the number of fire incidences during the first quarter of the year 2018 compared to its previous year (2017).

Accordingly, it is imperative to master the patterns and regularity of fire occurrences and make fire prevention countermeasures as soon as possible to reduce economic loss and human casualty. Therefore, it is of considerable significance to analyze the pattern and predict fire incidences using historical data (Zhanli, 2012; Zhang & Jiang, 2012; Ahn et al., 2015). One of the most commonly used methods found in literature in modeling time series data is the Box-Jenkins Method or the Autoregressive Integrated Moving Average (ARIMA) family of models. The Box-Jenkins method is efficient in capturing the processes by which the observations are generated. Thus, we use this method in the study. The study aims to a) analyze the characteristics of the historical data of fire incidence in Davao City, b) develop an appropriate model using the Box-Jenkins method for the fire incidence in Davao City, and c) forecast the monthly fire incidence in Davao City, using the appropriate model in b).

**Conceptual Framework**

In this study, we use the principle of modeling for decision making, which assumes that a model is an abstract representation of reality (Powell & Baker, 2009; Barker, 2005; Benjamin et al., 1998). Specifically, this principle has a framework that includes a) problem or some realities that are being investigated, b) an abstraction of the problem such as assumptions and data, c) the model which is built from the data, d) the output of the model that may provide insights to manage or solve the problem.

For the modeling phase, we use the framework for the time series analysis presented by Box and Jenkins (Montgomery et al., 2015), an iterative three-stage process of model identification, parameter estimation, and statistical diagnostic checking. To examine the forecasting performance of our model, another stage is added in the modeling phase, as suggested by other authors, which is forecast evaluation (Hyndman, 2001; Adhikari & Agrawal, 2013). A pre-processing stage is also introduced in this framework to check the data for any problems such as missing values and outliers (Kurbalija et al., 2010). The framework is summarized in the figure below.

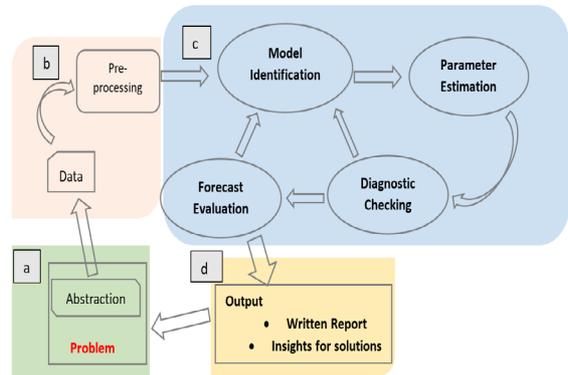

**Figure 1.** *Framework of the Study*

**2.0 Methods**

The datasets used in this study are obtained from the office of Bureau of Fire Protection – Region XI located at Alvarez Street corner Monteverde, Davao City. It is composed of 84 monthly fire incidence cases recorded from January 2012 to December 2018. The first 72 data points, from January 2012 to December 2017 have been used for model building (training set), while the next 12 data points, from January 2018 to December 2018 have been used for forecast evaluation (validation set). The complete data set is listed in Appendix A.

The Box-Jenkins method or the ARIMA family of models is used to analyze and forecast the monthly fire incidence in Davao City. The method includes three iterative stages: model-identification, model or parameter estimation, and diagnostic checking (Montgomery et al., 2015). Another stage is added, as suggested by other authors, which is forecast evaluation (Hyndman, 2001; Adhikari & Agrawal, 2013).

The model identification stage involves an examination of the characteristics of the time series. Characteristics of a non-stationary time series include the trend, seasonality, and cycle. Stationary time series possesses constant mean and constant variance, an essential requirement in modeling time series using the Box-Jenkins method. Transformation using differencing and Box-Cox is done when the original time series is non-stationary. Using the stationary series, tentative models are identified based on the characteristics and structure found in the autocorrelation plot (ACF) and partial autocorrelation plot (PACF). The Aikake Information Criterion (AIC) is used to identify the model subjected to parameter estimation and diagnostic checking.

In the model estimation stage, the parameters are estimated using conditional least squares and maximum likelihood. For robust prediction, the estimates of parameters must be statistically significant. If at least one of the estimates is not significant, we drop the model and go back to model identification by selecting the tentative model with the second lower AIC. The process is repeated until the condition of statistical significance is met.

The third stage is diagnostic checking or model diagnostics, where the adequacy of the model is checked through the residuals analysis. In this study, the Ljung-Box test is used to test a time series model's lack of fit. The null hypothesis of the test states that the model is fit for the data.

The final stage is forecast evaluation, which involves the computation of the one-step-ahead forecast and the calculation of the forecast error. The forecasting efficiency of the model is assessed using the forecast error. When the forecast error exhibits a white noise behavior based on the ACF and PACF of forecast errors, the model is considered efficient and useful.

Several studies applied ARIMA to build a forecasting model for diseases such as malaria (Perez & Ceballos, 2019), electricity prices (Contreras et al., 2003), wind speed (Kavasseri & Seetharaman, 2009), overseas US dollar remittances in the Philippines (Manayaga and Ceballos, 2019) and HIV cases in the Philippines (Tatoy and Ceballos, 2019). The general ARIMA model is given by

$$\Phi(B)(1-B)^d y_t = \delta + \Theta(B)\varepsilon_t$$

where $\delta$ is the constant term,
$\varepsilon_t$ is the white noise and
$$\Phi(B) = 1 - \sum_{i=1}^{p}\phi_i B^i$$
$$\Theta(B) = 1 - \sum_{j=1}^{q}\theta_j B^j$$

The R statistical software is used in all calculations in this study, including the generation of necessary plots. The following are the R packages used in this study: 'tseries' for testing the stationarity, 'astsa' for getting the numerical values and plots of ACF and PACF, 'forecast' for data transformation and fitting the ARIMA model, and 'lmtest' for getting the coefficient values of the selected model in model estimation. For the detailed R programming codes used in the analysis, please see Appendix B. The R programming codes are divided into the major procedural steps, namely, loading of R libraries, loading of data, model identification, model estimation, diagnostic checking, forecast evaluation and generation of forecast values.

**3.0 Results and Discussion**

In the Philippines, the National Fire Prevention Month is in March of every year under Proclamations No. 115-A and No. 360 signed by then-President Marcos in 1986. March is chosen because it is considered as the beginning of the hot season. Also, reports of the Fire Bureau show alarming fire incidences this month (Philippine Information Agency, 2019). To examine if there is a clustering of large fire incidences in the month of March, the time series plot of the monthly reported fire incidence in Davao City is presented in Figure 2. The highest reported number of fire incidence is 101 in March 2016. However, over the years, the plot did not show a clustering pattern in the month of March or, in other words, the highest fire incidence per year is not necessarily recorded in the month of March. The most massive occurrences of fire are found between the years 2015 and 2016. Most of these are due to unattended burned out candles in urban residential houses during brownouts. The brownouts are caused by the preventive maintenance work on major power plants and the reduced water level at hydropower sources caused by El Niño. One major fire incident took a Pastor's life in a church staff house in the Poblacion district of Davao City (Colina, 2015). In 2016, major fire incidence has been mostly in densely populated and congested urban residences. One major fire incidence has been in Bankerohan Public Market (National Disaster Risk Reduction and Management, 2017). There is no apparent trend or seasonality in the historical data of Fire incidence in Davao City. However, the series is nonstationary, and transformation is needed to stabilize the series.

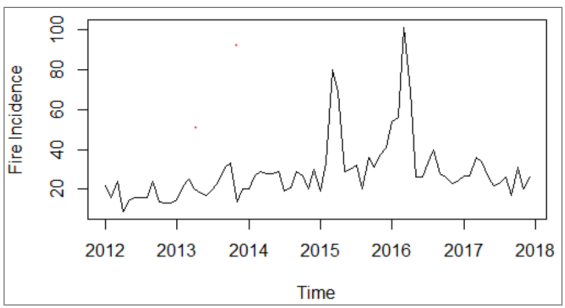

**Figure 2.** Time series plot of fire incidence in Davao City

### Model Identification

To stabilize the variance of the original time series, we implement the use of the Box-Cox transformation. Figure 3 shows the plot of the Box-Cox transformed series with λ=0. Based on the figure, the variance seems to improve, and it can be considered stable. However, the series remains nonstationary based on the observed changes in the series' mean levels over time.

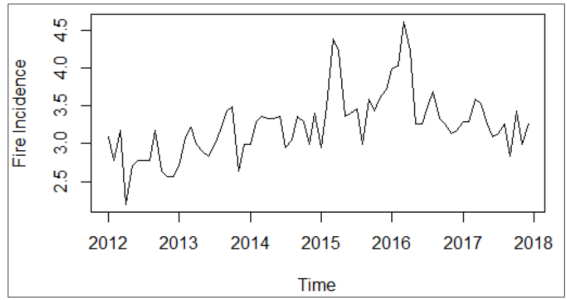

**Figure 3.** Time series plot of the transformed series

Since the transformed series is not yet stationary, we applied first differencing to the transformed series to address non-stationarity. The time series plot of the differenced series is given in Figure 4. The plot shows that the mean levels of the series have been stabilized using the method of differencing.

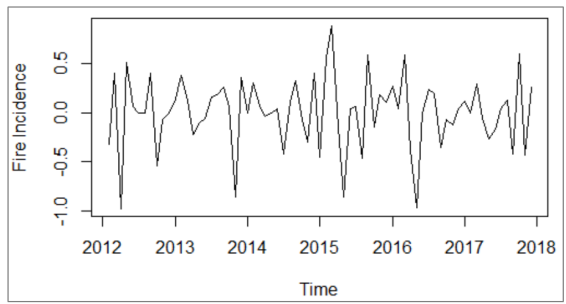

**Figure 4.** Time series plot of the differenced series

The Augmented Dicky Fuller test is performed as a formal test on the stationarity of the differenced series. The null hypothesis of the test states that the series is not stationary. Since the p-value (<0.01) is less than $\alpha$=0.05, the series is considered stationary, and the next step is to identify and list the tentative models. The autocorrelation plot (ACF) and partial autocorrelation plot (PACF) are used as the bases for selecting the tentative models (Keshvani, 2013).

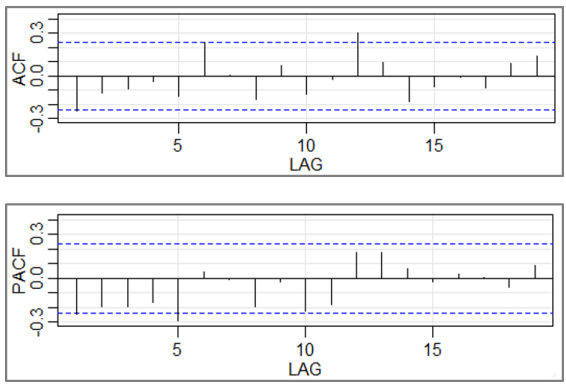

***Figure 5.*** *ACF and PACF Plots of the differenced series*

Figure 5 shows the ACF and PACF plots of the differenced series, respectively. Several interpretations can be made using the ACF and PACF plots. One is that the ACF cuts off at lag 1 (q=1) and PACF tails off (p=0), suggesting an ARIMA (0, 1, 1) model. Another interpretation is, the PACF cuts off at lag 1 (p=1), and the ACF tails off (q=0), suggesting an ARIMA (1, 1, 0) model. Additional interpretation will be, both ACF and PACF tails off, suggesting a model with both AR and MA terms or an ARIMA(p,d,q) where p and q can range from 1 to 3 and d=1. To identify the best ARIMA model among the tentative models for the fire incidence in Davao City, the Akaike Information Criterion (AIC) is used. Table 1 presents the tentative ARIMA models for the monthly fire incidence in Davao City, together with their respective AIC values. The best model in the list is ARIMA (1, 1, 1) since it has the smallest AIC value.

***Table 1.*** *List of tentative models with AIC values*

| Model | AIC value |
|---|---|
| ARIMA (0, 1, 1) | 52.84 |
| ARIMA (1, 1, 0) | 57.05 |
| ARIMA (1, 1, 1) | 48.09 |
| ARIMA (2, 1, 1) | 49.82 |
| ARIMA (3, 1, 1) | 51.12 |

### Model or Parameter Estimation

Table 2 shows the estimates, standard error, z-value, and p-value of the parameters of the model. The p-values of the autoregressive component or AR(1) and moving average component or MA(1) are both <0.01, which are less than $\alpha=0.05$ level of significance. Therefore, the estimates of the parameters are significantly different from zero. This is a good indication that the model is robust for prediction. The next step is to perform diagnostic checks on the model.

***Table 2.*** *Coefficient Estimates of ARIMA (1, 1, 1)*

| | Estimates | Std. Error | z-value | p-value |
|---|---|---|---|---|
| AR (1) | 0.400784 | 0.145218 | 2.7599 | <0.01 |
| MA (1) | -0.856217 | 0.076403 | -11.2066 | <0.01 |

### Diagnostic Checking

Residuals analysis is done to determine if the model meets the criteria for adequacy. See Appendix C: Residual Plots for reference. The plot of residuals vs. time does not show any strong pattern suggesting that the residuals are independent against time. The variance of the residuals is also constant. Hence, we can assume that the constancy of error variance is met. The residuals are near the theoretical line of normality, suggesting that the residuals are normally distributed. Lastly, most of the autocorrelations and partial autocorrelations are within acceptable limits. Thus, the residuals are considered white noise.

To formally test the model's lack of fit, the Ljung-Box test is performed and shown in Table 3. The null hypothesis of the Ljung-Box test states that the model does not exhibit a lack of fit. Since the p-value (0.7647) is higher than α=0.05, we cannot reject the null hypothesis that the model does not exhibit a lack of fit. Therefore, we can conclude that the model fits the data.

***Table 3.*** *Ljung-Box test for lack of fit of the model fit*

| Test Statistic | df | p-value |
|---|---|---|
| 6.5773 | 10 | 0.7647 |

***Forecast Evaluation***

The ARIMA(1,1,1) model is used together with the validation set (out-of-sample values) to obtain the one-step-ahead forecast value from January 2018 up to December 2018. The one-step-ahead forecast is then compared against the 12 out-of-sample actual values. Table 4 summarizes the actual values, one-step-ahead forecasted values, and forecast errors of the ARIMA (1,1,1) model.

There are no spikes in the ACF and PACF plots of the forecast errors indicating that the values are not significantly different from zero. Thus, the forecast errors behave like white noise. See Appendix D: Plots for forecast errors.

To formally test the normality of the forecast errors, the Shapiro-Wilk test is used. The null hypothesis of the Shapiro-Wilk test states that the forecast errors are normal. Since the p-value (0.3842) is higher than α=0.05, we cannot reject the null hypothesis that the error terms are normally distributed. Thus, the forecast errors are considered Gaussian white noise.

***Table 4.*** *Actual and one-step ahead forecasted values*

| Date | Actual Value | Forecasted Value | Forecast Error |
|---|---|---|---|
| Jan-2018 | 37 | 26 | 11 |
| Feb-2018 | 33 | 32 | 1 |
| Mar-2018 | 52 | 31 | 21 |
| Apr-2018 | 39 | 40 | -1 |
| May-2018 | 38 | 35 | 3 |
| Jun-2018 | 28 | 35 | -7 |
| Jul-2018 | 52 | 30 | 22 |
| Aug-2018 | 40 | 42 | -2 |
| Sep-2018 | 25 | 38 | -13 |
| Oct-2018 | 31 | 30 | 1 |
| Nov-2018 | 48 | 32 | 16 |
| Dec-2018 | 43 | 41 | 2 |

Table 5 shows the forecasted number of fire incidence in Davao City for the next 12 months (January 2019 up to December 2019) together with its lower and upper 95% confidence interval. It can be observed that all the forecasted values are within the specified 95% confidence intervals. Computing confidence intervals is an essential part of the forecasting process intended to indicate the likely uncertainty in point forecasts. Notice that the confidence interval is narrow for earlier dates or time and is wider for later dates or time.

Table 5 further shows that, on average, the forecasted fire incidence does not exhibit increasing patterns, consistent with our initial assessment that the observed fire incidence has no trend and seasonality. Also, there is no evidence suggesting that fire incidence will exceed the highest case observed in 2016, which is quite good news. However, some efforts to control the incidence of fire in Davao City must be put in place. Considering the behavior of the forecast values, there is no evidence that suggests cases will decrease in the long run.

*Table 5.* Forecasted values and their 95% confidence interval

| Date | Forecast | Lower 95% CI | Upper 95% CI |
|---|---|---|---|
| Jan-2019 | 39 | 21 | 75 |
| Feb-2019 | 37 | 19 | 79 |
| Mar-2019 | 36 | 18 | 80 |
| Apr-2019 | 38 | 17 | 82 |
| May-2019 | 39 | 17 | 83 |
| Jun-2019 | 37 | 17 | 84 |
| Jul-2019 | 36 | 16 | 85 |
| Aug-2019 | 35 | 16 | 86 |
| Sep-2019 | 36 | 16 | 88 |
| Oct-2019 | 37 | 16 | 89 |
| Nov-2019 | 37 | 15 | 90 |
| Dec-2019 | 37 | 15 | 91 |

**4.0 Conclusions**

This study aims to understand the characteristics or behavior of the cases of fire incidence in Davao City based on its available time-series data. Our analysis shows that the historical data of fire occurrences do not show trends and seasonality, which indicates that the number of fire occurrences in the city has not continued to increase or decrease over time. A consistent decrease of an adverse event like fire incidence is an idealistic aim because it would mean that the cases would eventually reach zero over time. On the other hand, a consistent increase in fire incidence poses a severe problem and would require an immediate action and response to mitigate the grave loss of life and property. Although the fire incidence in Davao City may not go up consistently over time, it may also not die down naturally. Thus, it is imperative to develop plans, programs, and policies to mitigate the incidence of fire or to prevent it from occurring.

To aid in the process of decision making or development of programs, a forecasting model is developed in this study to measure the expected cases of fire incidence. The forecasting model takes into account the behavior of the observed fire incidence, including the fact that there are no seasonal patterns in the data. The absence of a seasonal pattern implies no specific month or period where the surge of cases is observed. The surge of fire incidence may occur at any time of the year. Therefore, fire prevention activities should be done all year round and not just during the fire prevention month.

The best model to forecast the monthly reported fire incidence in Davao City is ARIMA (1,1,1). This model has undergone stages of validation to assess its robustness and validity in generating the forecast. Generally, the model is useful for forecasting purposes. However, since there are few significant spikes in ACF and PACF of residuals, this model shall be used cautiously. Based on the generated forecasts, fire incidence cases in the city range between 36 to 39 cases per month on average. The cases can be as low as 15 per month, or it can be as high as 91 cases per month, considering the 95% confidence interval.

**Recommendations and Insights for solutions**

The following are the recommendations based on the findings and results of our analysis.

1. Contrary to our assumption, fire incidence in Davao City do not follow seasonal patterns. Our analysis shows

that cases do not surge during hot season, and it also do not drop in the wet season. For this reason, fire prevention programs should not only be intensified or campaigned during the National Fire Prevention Month, which is March of every year. There should be intensification of the campaign all year round.
2. A higher incidence of fire occurs in residential urban areas with high population density. These are mostly congested areas closely located at the heart of the city. The local government unit may propose improvement plans to decongest these areas to prevent substantial damages when fire incidence occurs.
3. There should be close coordination between the electricity provider and the Bureau of Fire Prevention XI, specifically when preventive maintenance work is done so that BFP can plan, prepare, and control possible fire incidence during the work.

***Appendix A.*** *Data set of Fire Incidence*

Training Set

| | | | | | | | |
|---|---|---|---|---|---|---|---|
| Jan-12 | 22 | Jul-13 | 20 | Jan-15 | 19 | Jul-16 | 33 |
| Feb-12 | 16 | Aug-13 | 24 | Feb-15 | 33 | Aug-16 | 40 |
| Mar-12 | 24 | Sep-13 | 31 | Mar-15 | 80 | Sep-16 | 28 |
| Apr-12 | 9 | Oct-13 | 33 | Apr-15 | 69 | Oct-16 | 26 |
| May-12 | 15 | Nov-13 | 14 | May-15 | 29 | Nov-16 | 23 |
| Jun-12 | 16 | Dec-13 | 20 | Jun-15 | 30 | Dec-16 | 24 |
| Jul-12 | 16 | Jan-14 | 20 | Jul-15 | 32 | Jan-17 | 27 |
| Aug-12 | 16 | Feb-14 | 27 | Aug-15 | 20 | Feb-17 | 27 |
| Sep-12 | 24 | Mar-14 | 29 | Sep-15 | 36 | Mar-17 | 36 |
| Oct-12 | 14 | Apr-14 | 28 | Oct-15 | 31 | Apr-17 | 34 |
| Nov-12 | 13 | May-14 | 28 | Nov-15 | 37 | May-17 | 26 |
| Dec-12 | 13 | Jun-14 | 29 | Dec-15 | 41 | Jun-17 | 22 |
| Jan-13 | 15 | Jul-14 | 19 | Jan-16 | 54 | Jul-17 | 23 |
| Feb-13 | 22 | Aug-14 | 21 | Feb-16 | 56 | Aug-17 | 26 |
| Mar-13 | 25 | Sep-14 | 29 | Mar-16 | 101 | Sep-17 | 17 |
| Apr-13 | 20 | Oct-14 | 27 | Apr-16 | 69 | Oct-17 | 31 |
| May-13 | 18 | Nov-14 | 20 | May-16 | 26 | Nov-17 | 20 |
| Jun-13 | 17 | Dec-14 | 30 | Jun-16 | 26 | Dec-17 | 26 |

Validation Set

| | | | |
|---|---|---|---|
| Jan-18 | 37 | Jul-18 | 52 |
| Feb-18 | 33 | Aug-18 | 40 |
| Mar-18 | 52 | Sep-18 | 25 |
| Apr-18 | 39 | Oct-18 | 31 |
| May-18 | 38 | Nov-18 | 48 |
| Jun-18 | 28 | Dec-18 | 43 |

***Appendix B.*** *R Programming Codes*

Loading the required Libraries

```
> library(astsa)
> library(forecast)
> library(lmtest)
> library(tseries)
```

Loading the data into R

```
> finaldata=ts(THESIS_data_72_first_,frequency=12,start=c(2012,1))
```

Model Identification

```
> plot(finaldata)
> Acf(finaldata,main="")

#ADF test for stationarity of the original series
> adf.test(finaldata)

#Box-Cox transformation
> newdata=BoxCox(finaldata,0)
> plot(newdata)
> acf2(newdata)

#ADF test for stationarity of the transformed series
> adf.test(newdata)

#Applying first differencing to the transformed series
> d1=diff(newdata,1)
> plot(d1)
> acf2(d1,main="")
#ADF test for stationarity of first difference of the transformed series
> adf.test(d1)

#Tentative models
> model1=Arima(finaldata,order=c(0,1,1),lambda=0)
> model2=Arima(finaldata,order=c(1,1,0),lambda=0)
> model3=Arima(finaldata,order=c(1,1,1),lambda=0)
> model4=Arima(finaldata,order=c(2,1,1),lambda=0)
> model5=Arima(finaldata,order=c(3,1,1),lambda=0)
> model1$aic
> model2$aic
> model3$aic
> model4$aic
> model5$aic
```

Model Estimation

```
> model=model3
> coeftest(model)
```

Diagnostic Checking

```
#Residuals analysis
> plot(model$residuals,type="p",ylab="Residuals")
> abline(0,0)
> plot(asnumeric(model$fitted),model$residuals,xlab="Fitted",ylab
="Residuals")
> abline(0,0)
> qqnorm(model$residuals)
> qqline(model$residuals)

#ACF and PACF values and plots of residuals
> acf2(model$residuals,main="")

#Shapiro-Wilk test of normality for residuals
> shapiro.test(model$residuals)

#Ljung-Box test for residuals
> Box.test(model$residuals,typ="Lj",lag=10)
```

Forecast Evaluation

```
> validation=read_excel("THESIS_data_12_last_.xlsx")
> validation=ts(THESIS_data_12_last_)
> newmodel=Arima(c(finaldata,validation),model=model,lambda=0)

#One-step ahead forecasted values
> onestep=fitted(newmodel)[73:84]

#Forecast Errors
> fe=validation-onestep

#ACF and PACF values and plots of forecast errors
> acf2(fe,main="")
> qqnorm(fe)
> qqline(fe)

#Shapiro-Wilk test of normality for forecast errors
> shapiro.test(fe)
```

Forecast Evaluation *(continued...)*

```
#Arrange
> evaluation=cbind(validation,onestep,fe)
> colnames(evaluation)=c("Actual","Forecast","Forecast Error")
> evaluation
```

Generate forecasts

```
> fore=forecast(c(finaldata,validation),model=model,h=12)
> print(fore)
```

***Appendix C.*** *Residual Plots*

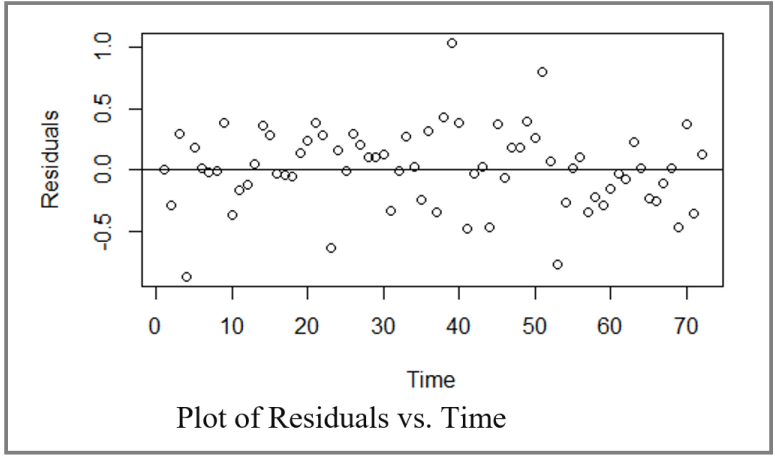

Plot of Residuals vs. Time

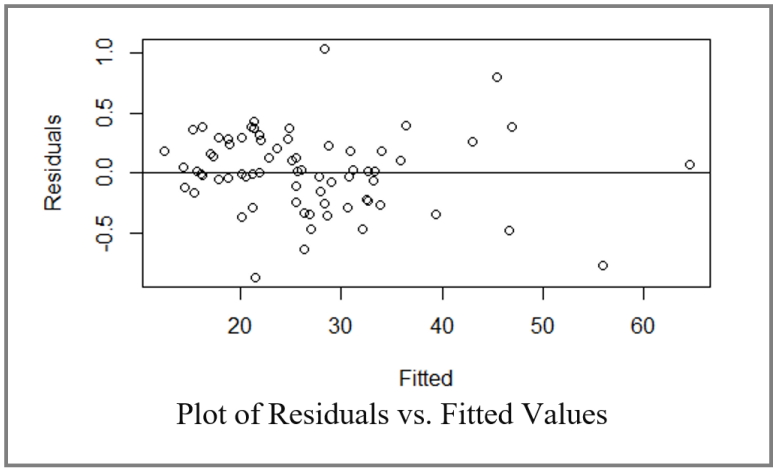

Plot of Residuals vs. Fitted Values

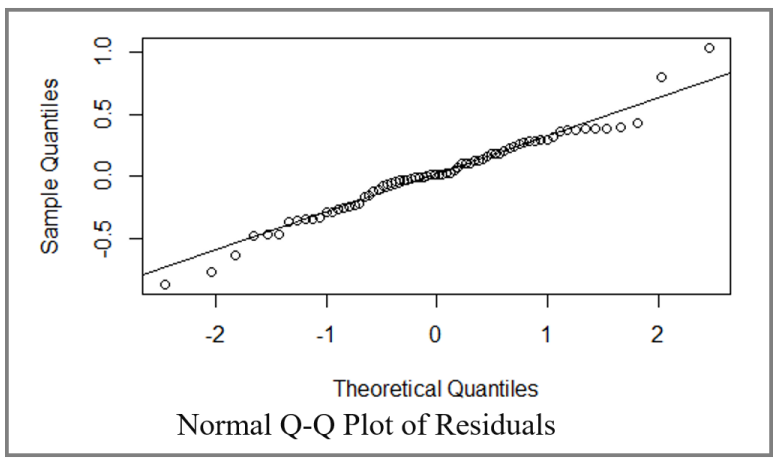

Normal Q-Q Plot of Residuals

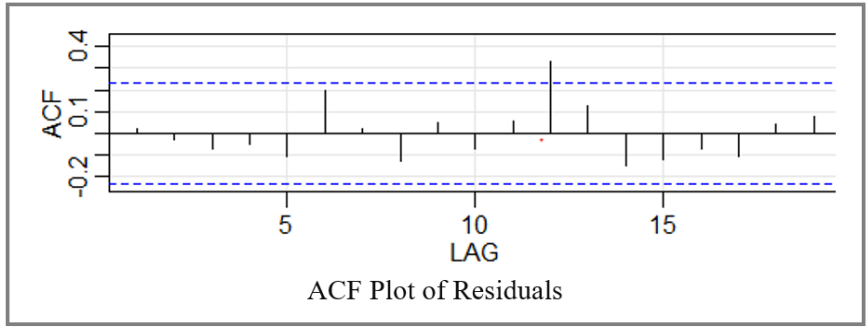
ACF Plot of Residuals

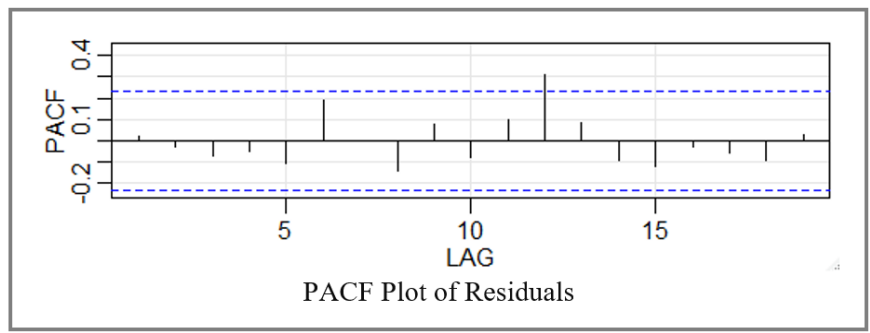
PACF Plot of Residuals

**Appendix D.** *Plots of forecast errors*

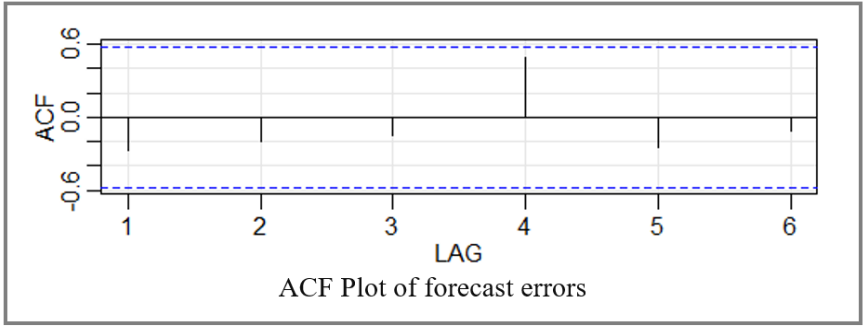

ACF Plot of forecast errors

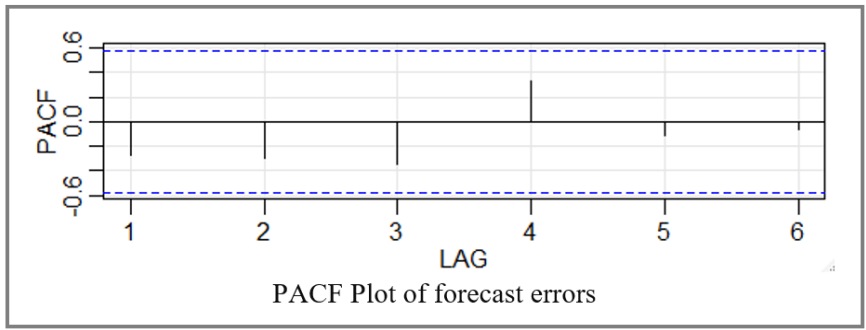

PACF Plot of forecast errors